# Multiscale simulation of atomic structure in the vicinity of nanovoids and evaluation of the shifting rates of the void surface elements in bcc iron and aluminum

A.V. Nazarov[1,2], A.P. Melnikov[1], A.A. Mikheev[3], I.V. Ershova[1]

[1] National Research Nuclear University MEPhI, (Moscow Engineering Physics Institute), 31, Kashirskoe shosse 115409, Moscow, Russia.
[2] Institute for Theoretical and Experimental Physics named by A.I. Alikhanov of NRC "Kurchatov Institute", 25 Bolshaya Cheremushkinskaya str.,117218, Moscow, Russia
[3] The Kosygin State University of Russia, 33 Sadovnicheskaya str., 117997, Moscow, Russia.

**E-mail:** avn46@mail.ru

**Abstract**. We simulate structure in the vicinity of different size nanovoids using a new variant of the Molecular Statics, wherein atomic structure in the vicinity of nanovoids and the parameters that define the displacements of atoms placed in elastic continuum around main computation cell are determined in a self-consistent manner. Then we model structure of the surface of different crystallographic planes. Next, the kinetic equations for shifting rate of void surface elements located normal to the void surface are obtained. These equations take into account the dependence of the vacancy flux on strain and the surface energies of the crystallographic planes. In the next section, we apply the obtained expressions and simulation results to calculate the shifting rates of void surface elements for different crystallographic directions. Rates of displacements in different crystallographic directions for bcc and fcc metals differ significantly, and for the <100> direction they are greatly lower than for other directions.

## 1. Introduction

It is known that voids form and grow in materials under irradiation [1-5]. Moreover, in many cases, initially spherically symmetric voids acquire a cuboid shape over time. [3, 4, 6]. According to the works [3, 7], several factors influence the change in the void shape upon irradiation: surface energy anisotropy for different crystallographic planes, preferred adsorption of atoms on certain crystallographic surfaces and vacancy diffusion fluxes anisotropy in the vicinity voids.

Ordinarily the elastic fields around the voids are not taken into account in models that predicting the material's behavior under irradiation [1]. However, as shown previously [8-14], these fields affect the flow of defects and in the case of nanovoids have a significant effect on vacancy fluxes [11,12,14], and, consequently, on the growth and dissolution rates of voids. According to the results of [10] the X-axis component of the vacancy flux in the zero approximation of the elastic field effect is described by the following equation:

$$J_x = -D_V \left[ \frac{\partial c}{\partial x} - c \frac{K^V}{kT} \frac{\partial Sp\varepsilon}{\partial x} \right], \qquad (1)$$

where $J_x$ is the vacancy diffusion flux in the direction of X-axis, $c$ is the vacancy concentration, $D_V$ is the vacancy diffusion coefficient in an ideal system, $K^V$ is the coefficient that determines the elastic field contribution to the vacancy flux [10], $\varepsilon$ is the strain tensor, $Sp\varepsilon$ is the strain tensor trace, $k$ is the Botzmann constant, $T$ is the temperature.

As a rule displacement fields in the vicinity of defects are described by equations of the elasticity theory. The solution of the equations for an isolated void of a spherical shape with radius $R$ has the form [15]:

$$u_i = C_1 \frac{x_i}{r^3}, \qquad (2)$$

where $u_i$ is the atom displacement from the initial position in the $x_i$ direction, $r$ is the distance from the center of the void, and $C_1$ determined by both the characteristics of the material and the void radius:

$$C_1 = -\frac{1+\nu}{2E} \gamma R^2, \qquad (3)$$

where $R$ is the radius of the spherical void, $\gamma$ is surface energy, $\nu$ is the Poisson's ratio, $E$ is the Young's modulus.

The components of the strain tensor are determined by the equation:

$$\varepsilon_{ij} = \frac{1}{2}(\frac{\partial u_i}{\partial x_j} + \frac{\partial u_j}{\partial x_i}). \tag{4}$$

When differentiate the displacement (Equation (2)) we get the equation for the diagonal components of the strain tensor:

$$\varepsilon_{ii} = C_1(\frac{1}{r^3} - \frac{3x_i^2}{r^5}), \tag{5}$$

and the trace of the strain tensor is a constant, and its derivative is zero:

$$\nabla Sp\varepsilon = 0. \tag{6}$$

By that means, elastic fields have no effect on diffusion fluxes and pore growth rates., and in that connection voids are often called neutral sinks [1]. In our work, we use the Modified Molecular Static [16-19] for modeling atomic structure in the vicinity of the voids and finding atom displacements [12,14,20]. This approach takes into account the discreteness of the structure and its anisotropy. As a result, the atoms' displacements for different crystallographic directions differ significantly, and the trace of the strain tensor is different from the constant and its gradient is not zero. Therefore, an additional term appears in the equations for vacancy fluxes, and this term differ for various crystallographic directions.

The model is described in detail in [18-20], there are also some results for α-Fe. The results of modeling the structure are used to calculate in the next section the dependences of the shifting rate of the void surface elements on the oversaturation in α-Fe and aluminum over a wide temperature range. Moreover, in our works [12,14 ] it was shown that these dependences for crystallographic directions of the type <100>, <110>, <111> differ significantly, and, thus, the anisotropy of the elastic field in the vicinity of the void can be the cause of a change in the shape of the initially spherical voids.

Now we calculate the surface energy for different crystallographic planes by the MS and take into account the influence of the effect on the shifting rate of the void surface elements.

This makes it possible to compare the effect of these two factors on vacancy fluxes and shape changes of the voids in metals.

## 2. Method

Atoms surround a spherical void with radius $R$, the positions of atoms are determined by the vector $r$. The system is divided into two zones – main computational cell (zone I, $R_S$) and elastic medium in which atoms are immersed (zone II). The zone structure is shown on Figure 1.

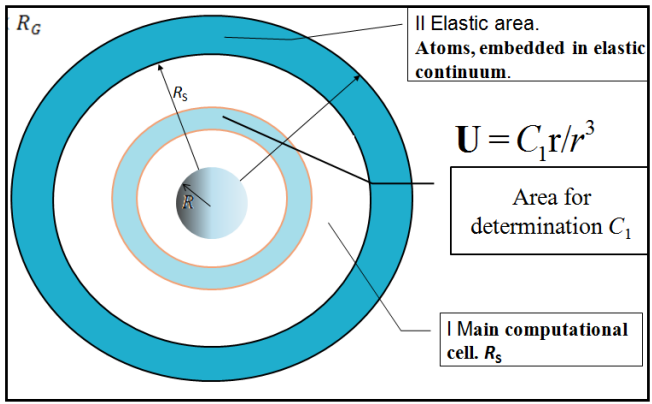

**Fig. 1** Scheme of the computational cell in the vicinity of the void.

The atomic coordinates of the first zone are calculated using the variational procedure of the MS method. The atoms surrounding the main computational cell are located in an elastic medium, and their displacements are determined based on solutions of the elasticity theory equations (2):

An important feature of the model is a self-consistent iterative procedure for calculating the atoms' positions in the main computational cell and calculating the constant $C_1$, which determines the displacements in the elastic zone. The constant $C_1$ is calculated using Eq. (2) based on the results of modeling atomic displacements in the spherical layer located approximately in the middle between the defect and the boundary of the main computational cell. The simulation results indicate a stable convergence of the iterative procedure. The model features are described in [12,20].

The model allows one to obtain a structure in the vicinity of voids [12,14,20]. The next section presents the simulation results for voids of some sizes in α-Fe and Al that are further used in calculating the shifting rate of void surface elements.

## 3. Simulation results of atomic structure in vicinity of nanovoids

Figures 2 show the dependences of atomic displacements for crystallographic directions of the type <100>, <110>, <111> on the distance to the void center in α-Fe and aluminum. Calculations of atomic displacements based on the solution of the isotropic theory of elasticity equation are also given for comparison for voids of the same size. These solutions have spherical symmetry and are independent of angles.

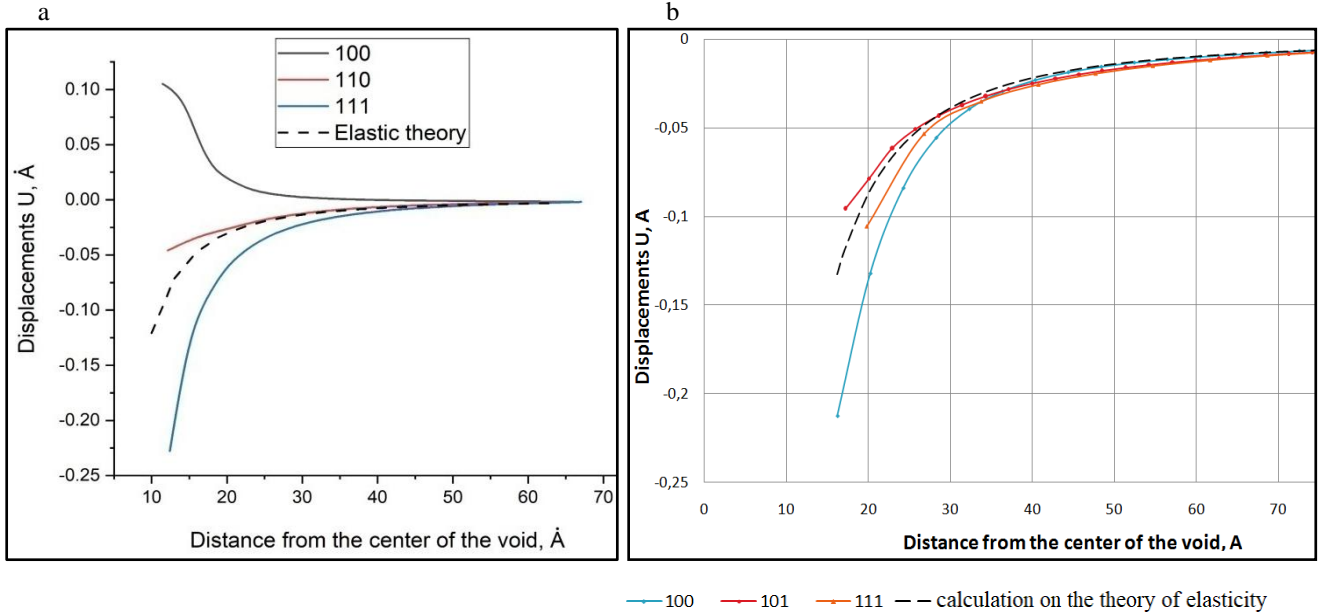

**Fig. 2** The atoms' displacements for various directions in α-Fe (a) for nanovoid (R = 12.01 Å) and aluminum (b) for nanovoid (R = 16.91 Å)

In contrast to the predictions of the isotropic theory of elasticity, atoms' displacements in different crystallographic directions for α-Fe differ significantly. For directions of type <100>, the displacements near the voids are positive. The maximum atoms' displacements from the site of the ideal lattice are observed in the direction <111>. It should be noted that according to the simulation results for voids of different sizes (up to 20 Å [20]), the dependences have a similar form. However, in contrast to bcc iron, for aluminum (Figure. 2b), the displacements of atoms for different crystallographic directions are negative and close in magnitude.

From the analysis of the results it follows that the trace of the strain tensor is not zero, and the equations for vacancy fluxes, and consequently the kinetic equations for the void growth rate must contain additional terms due to the elastic field:

$$\nabla Sp\varepsilon \neq 0. \tag{8}$$

Therefore, in the next section, we present the derivation of equations for the shifting rate of void surface elements based on the results of the original theoretical approach developed by us in recent years [8-10] that makes it possible to describe diffusion fluxes under stress. According to the results of [10], the component of the vacancy flux in the X direction in the zeroth approximation with respect to strain has the form Eq. (1).

## 4. Shifting rate of the void surface elements for the different crystallographic directions

It can be shown in the same way as it was done in [21] that the movement of the shifting rate of void surface element in a certain direction is determined by the equation:

$$\frac{dR}{dt} = -\left(\vec{n}, \vec{j}\right), \tag{9}$$

where $R$ is the radius of the void, $n$ is the normal to the void surface, $j$ is vacancy flux density to the void surface.

To evaluate the influence of the elastic field on the vacancy flux in the region near the voids and obtain analytical solutions, the method of successive approximations is used. And as a first approximation, similar to [22], we choose the solution of the diffusion equation for the vacancy concentration in that the influence of the field is not taken into account. Then the change rate of the void radius $R$ is described by equation [22]:

$$\frac{dR}{dt} = C_{eq}D_V\left[\frac{R_G}{R_G - R}R^{-1}\left(\Delta + 1 - \exp\left(\frac{2\gamma V^f}{kTR}\right)\right)\right], \quad (10)$$

where $c_{eq}$ is the equilibrium vacancy concentration for the flat surface, $V^f$ is the vacancy formation volume, $\Delta = \frac{c_m - c_{eq}}{c_{eq}}$ is the oversaturation of vacancies, $c_m = c(R_G)$, where $R_G$ is half the average distance between voids.

Transformations for getting the equations for the shifting rate of the void surface elements for various crystallographic directions taking into account the field of elastic strains are given in [12]. Now we have taken into account the different values of the surface energy for different crystallographic planes. Here, we restrict ourselves to providing final equations for the shifting rate of the void surface elements provided that the void density is low, $R \ll R_G$:

$$\frac{dR_{100}}{dt} = c_{eq}D_V\left[\frac{1}{R}\left(\Delta + 1 - exp\left(\frac{2\gamma_{100}V^f}{kTR}\right)\right) - \frac{K^V}{kT}exp\left(\frac{2\gamma_{100}V^f}{kTR}\right)\frac{\partial Sp\varepsilon}{\partial x}\right], \quad (11)$$

$$\frac{dR_{110}}{dt} = \frac{c_{eq}D_V}{\sqrt{2}}\left[\frac{\sqrt{2}}{R}\left(\Delta + 1 - exp\left(\frac{2\gamma_{110}V^f}{kTR}\right)\right) - \frac{K^V}{kT}exp\left(\frac{2\gamma_{110}V^f}{kTR}\right)\left(\frac{\partial Sp\varepsilon}{\partial x} + \frac{\partial Sp\varepsilon}{\partial y}\right)\right], \quad (12)$$

$$\frac{dR_{111}}{dt} = \frac{c_{eq}D_V}{\sqrt{3}}\left[\frac{\sqrt{3}}{R}\left(\Delta + 1 - exp\left(\frac{2\gamma_{111}V^f}{kTR}\right)\right) - -\frac{K^V}{kT}exp\left(\frac{2\gamma_{111}V^f}{kTR}\right)\left(\frac{\partial Sp\varepsilon}{\partial x} + \frac{\partial Sp\varepsilon}{\partial y} + \frac{\partial Sp\varepsilon}{\partial z}\right). \quad (13)$$

were $\gamma_{100}$ is the surface energy of the plane <100>, $\gamma_{110}$ is the surface energy of the plane <110>, $\gamma_{111}$ is the surface energy of the plane <111>.

## 5. Results simulation and discussion

The components of the strain tensor for each of the crystallographic directions, the trace of the strain tensor, and its derivatives to the corresponding coordinates are calculated based on the approximated simulation results of atom displacements near the void surface. After that we can calculate the shifting rate of the void surface elements using Eqs. (11-13). It was done for various oversaturations in a wide temperature range. Figures 3 show the normalized dependences of the mentioned void growth rates for α-Fe at the oversaturation $\Delta = 20$.

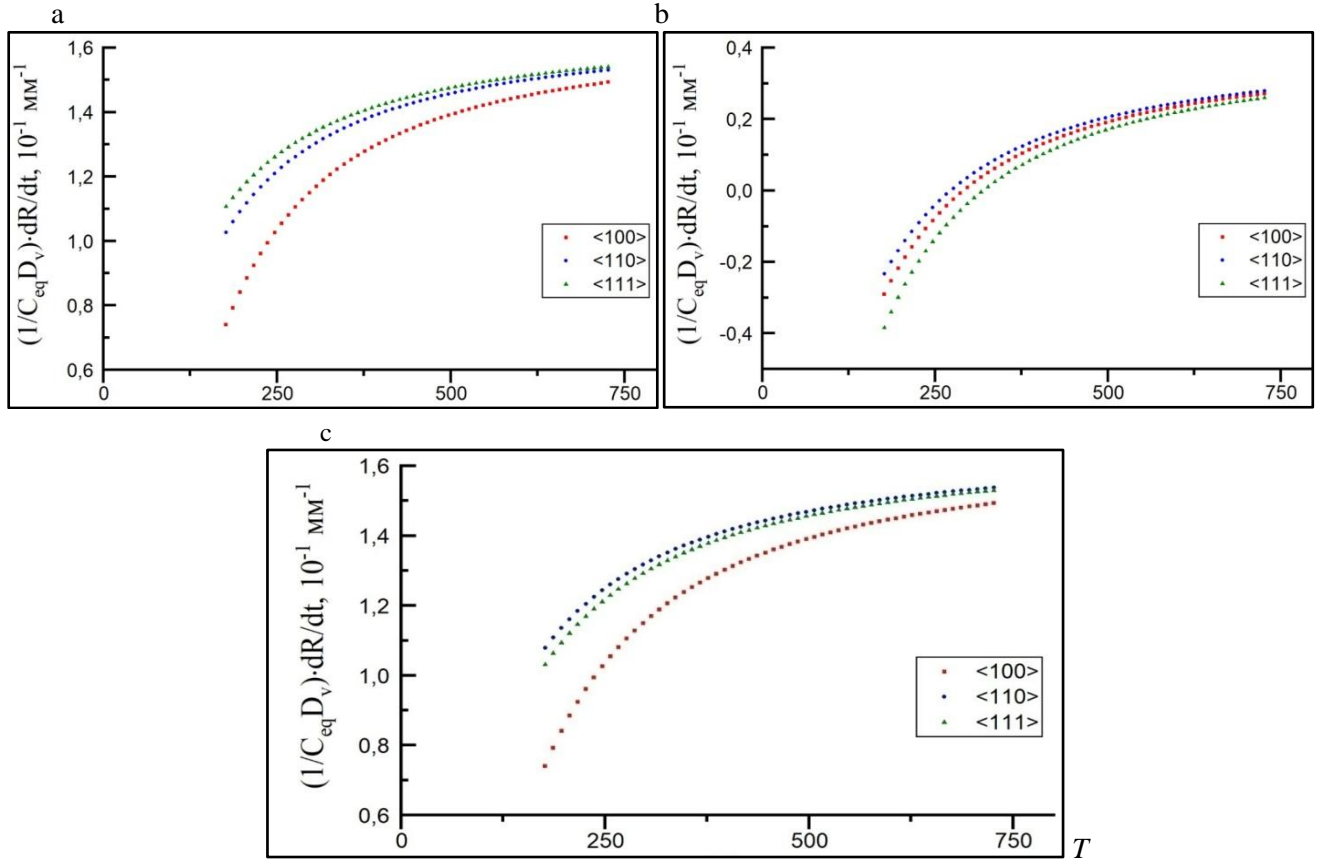

**Fig.** 3 Void growth rates for different directions, bcc Fe, R = 12.01 Å, Δ=20). *a* - taking into account the anisotropy of the elastic field, *b* - considering only the difference in surface energies for planes, *c* - taking into account the combined effect of both factors

In α-Fe the void growth rate significantly slows down in the directions of the <100> type due to the negative contribution of the strain. For directions of the <110> and <111> types the growth rates are quite close (Figure 3a). The influence of the difference in surface energies is small (Figure 3b). Figures 4 show the normalized dependences of the void growth rates for aluminium.

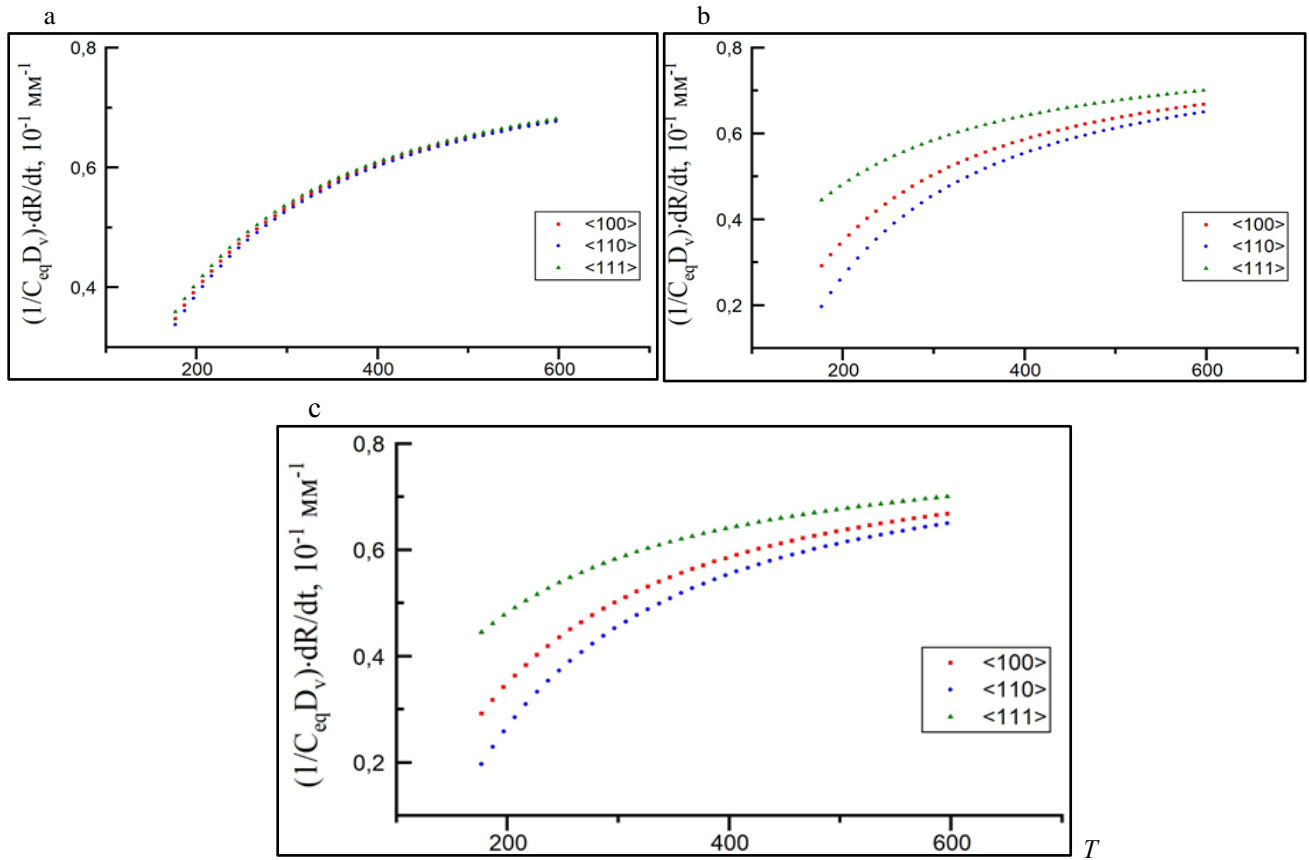

**Fig. 4** Void growth rates for different directions, fcc Al, R = 12.7 Å, Δ=10). a - taking into account the anisotropy of the elastic field, b- considering only the difference in surface energies for planes, c - taking into account the combined effect of both factors

As can be seen from Figures 4a–4c, the contribution of the elastic displacement field in fcc Al is much smaller compared to the influence of the difference in surface energies, and the main factor affecting the void growth kinetics is the difference in surface energies. In aluminum the void growth rate significantly slows down in the directions of the <110> type.

Thus, for α-Fe, main reason that influence the change in the shape of the initially spherical voids is the vacancy flux anisotropy due to the asymmetry of atomic displacements in the vicinity of voids. and due to the slowing down of the shifting rate of the void surface elements along the directions of the <100> type, the void shape will gradually be faceted by {100} planes. For aluminum, the main reason for the change in the shape of the voids is the difference in surface energies for crystallographic planes of different types.

## 6. Conclusion

- New model is devoted for determining atomic structure in the vicinity of nanovoids and calculations of the shifting rates of the void surface elements in bcc and fcc metals.
- The atom displacements near nanovoids are obtained that differ sufficiently from the elasticity theory solution.
- In addition, the displacement significantly different for variant crystallographic directions, and these differences are particularly large in metals with bcc structure.
- Calculations of surface energy are performed by using MS for some planes of bcc Fe and fcc Al.
- The calculations of the shifting rates of the void surface elements in bcc Fe and fcc Al are executed by using the results of atomic simulations.
- It is shown that the main reason for the anisotropy of the void growth rate in bcc iron is a significant difference in the atom displacements along different crystallographic directions, while in aluminum this reason is the significant differences in the surface energies of crystallographic planes.
- These results allow us to explain the faceted shape of the voids in the metals under conditions of substantial supersaturation with vacancies.

**Acknowledgements**

Authors would like to acknowledge the financial support of the National Research Nuclear University MEPhI Academic Excellence Project (Contract No. 02.a03.21.0005).